\documentclass[letterpaper, 12pt]{article}[2000/05/19]
\usepackage[english]{babel}
\usepackage{amsfonts,amsmath,amssymb,amsthm,latexsym,amscd,mathrsfs}
\usepackage{ifthen,cite}
\usepackage[bookmarksnumbered=true]{hyperref}
\hypersetup{pdfpagetransition={Split}}

\newcommand{\evenhead}{Author \ name}
\newcommand{\oddhead}{Article \ name}
\newcommand{\theArticleName}{Article \ name}

\newcommand{\FirstPageHeading}[1]{\thispagestyle{empty}%
\noindent\raisebox{0pt}[0pt][0pt]{\makebox[\textwidth]{\protect\footnotesize \sf }}\par}

\newcommand{\ArticleName}[1]{\renewcommand{\theArticleName}{#1}\vspace{-2mm}\par\noindent {\LARGE\bf  #1\par}}
\newcommand{\Author}[1]{\vspace{5mm}\par\noindent {\Large  #1\par} \par\vspace{2mm}\par}
\newcommand{\Address}[1]{\vspace{2mm}\par\noindent {\it #1} \par}
\newcommand{\Email}[1]{\ifthenelse{\equal{#1}{}}{}{\par\noindent {\rm E-mail: }{\it  #1} \par}}
\newcommand{\URLaddress}[1]{\ifthenelse{\equal{#1}{}}{}{\par\noindent {\rm URL: }{\tt  #1} \par}}
\newcommand{\EmailD}[1]{\ifthenelse{\equal{#1}{}}{}{\par\noindent {$\phantom{\dag}$~\rm E-mail: }{\it  #1} \par}}
\newcommand{\URLaddressD}[1]{\ifthenelse{\equal{#1}{}}{}{\par\noindent {$\phantom{\dag}$~\rm URL: }{\tt  #1} \par}}

\newcommand{\Abstract}[1]{\vspace{6mm}\par\noindent\hspace*{10mm}
\parbox{140mm}{\small {\bf Abstract.} #1}\par}
\newcommand{\Keywords}[1]{\vspace{3mm}\par\noindent\hspace*{10mm}
\parbox{140mm}{\small {\bf Key words:} \rm #1}\par}
\newcommand{\Classification}[1]{\vspace{3mm}\par\noindent\hspace*{10mm}
\parbox{140mm}{\small {\it 2010 Mathematics Subject Classification:} \rm #1}\vspace{3mm}\par}
\newcommand{\ShortArticleName}[1]{\renewcommand{\oddhead}{#1}}
\newcommand{\AuthorNameForHeading}[1]{\renewcommand{\evenhead}{#1}}

\long\def\@makecaption#1#2{
  \sbox\@tempboxa{\small \textbf{#1.}\ \ #2}%
  \ifdim \wd\@tempboxa >\hsize
    {\small \textbf{#1.}\ \ #2}\par \else
    \global \@minipagefalse
    \hb@xt@\hsize{\hfil\box\@tempboxa\hfil}%
  \fi \vskip\belowcaptionskip}

\def\numberwithin#1#2{\@ifundefined{c@#1}{\@nocounterr{#1}}{%
  \@ifundefined{c@#2}{\@nocnterr{#2}}{%
  \@addtoreset{#1}{#2}%
  \toks@\@xp\@xp\@xp{\csname the#1\endcsname}%
  \@xp\xdef\csname the#1\endcsname
    {\@xp\@nx\csname the#2\endcsname.\the\toks@}}}}
\def\E^#1{{\buildrel #1 \over\vee}}
\newtheorem{theorem}{Theorem}

{\theoremstyle{definition}

}

\begin{document}

\FirstPageHeading{V.I. Gerasimenko}

\ShortArticleName{Dynamics of correlations}

\AuthorNameForHeading{V.I. Gerasimenko}

\ArticleName{On Dynamics of Correlations in Quantum\\ Many-Particle Systems}
\Author{V.I. Gerasimenko\let\thefootnote\relax\footnote{E-mail: \emph{gerasym@imath.kiev.ua}}
}

\Address{Institute of Mathematics of NAS of Ukraine,\\
    3, Tereshchenkivs'ka Str.,\\
    01601, Kyiv, Ukraine}

\bigskip

\Abstract{
The paper deals with the problem of the rigorous description of the evolution of states of
large particle quantum systems by means of correlation operators. A nonperturbative solution
of the Cauchy problem of the hierarchy of nonlinear evolution equations for a sequence of
marginal correlation operators is constructed. For initial states specified in terms of a
one-particle density operator and correlation operators we also develop an approach to
the description of the processes of the creation and the propagation of correlations within
the framework of a one-particle density operator. Moreover, the mean field asymptotic
behavior of constructed marginal correlation operators is established.
}

\medskip


\vspace{2pc}
\Classification{35Q40; 82C40; 82C10.}

\makeatletter
\renewcommand{\@evenhead}{
\hspace*{-3pt}\raisebox{-15pt}[\headheight][0pt]{\vbox{\hbox to \textwidth {\thepage \hfil \evenhead}\vskip4pt \hrule}}}
\renewcommand{\@oddhead}{
\hspace*{-3pt}\raisebox{-15pt}[\headheight][0pt]{\vbox{\hbox to \textwidth {\oddhead \hfil \thepage}\vskip4pt\hrule}}}
\renewcommand{\@evenfoot}{}
\renewcommand{\@oddfoot}{}
\makeatother

\newpage
\vphantom{math}

\protect\tableofcontents
\vspace{0.5cm}

\section{Introduction: on the propagation of initial correlations}
In papers \cite{G14},\cite{GTsm} it was developed two new approaches to the description of the
propagation of initial correlations of large particle quantum systems in the mean field scaling
limit. We note that initial states specified by correlations are typical for the condensed
states of particle systems in contrast to their gaseous state \cite{BQ},\cite{S-R}.

In paper \cite{G14} the property of the propagation of initial correlations was proved within
the framework of the description of the evolution by means of marginal observables and in paper
\cite{GTsm} this property was established by another method in terms of a one-particle (marginal)
density operator governed by the generalized quantum kinetic equation \cite{GT10}. It was proved
that for a system of a non-fixed (i.e., arbitrary but finite) number of identical particles, obeying
the Maxwell -- Boltzmann statistics, in case of initial states specified by a sequence of the following
marginal correlation operators (we use notations accepted in review \cite{G12})
\begin{eqnarray*}\label{lins}
   &&g^{(c)}=\big(I,f_1^{0}(1),g_{2}(1,2)\prod_{i=1}^{2}f_1^{0}(i),\ldots,
        g_{n}(1,\ldots,n)\prod_{i=1}^{n}f_1^{0}(i),\ldots\big),
\end{eqnarray*}
in the mean field scaling limit the evolution of all possible correlations is described by the
following sequence of the limit marginal correlation operators
\begin{eqnarray}\label{pc}
   &&\hskip-9mm g_n(t,1,\ldots,n)=
       \prod_{i_1=1}^{n}\mathcal{G}_{1}^{\ast}(t,i_1)g_{n}(1,\ldots,n)
       \prod_{i_2=1}^{n}(\mathcal{G}_{1}^{\ast})^{-1}(t,i_2)\prod\limits_{j=1}^{n}f_{1}(t,j),
       \quad n\geq2,
\end{eqnarray}
where the one-particle density operator $f_{1}(t)$ is a solution of the Cauchy
problem of the quantum Vlasov-type kinetic equation with initial correlations:
\begin{eqnarray}\label{Vkec}
  &&\hskip-5mm\frac{\partial}{\partial t}f_{1}(t,1)=\mathcal{N}^{\ast}(1)f_{1}(t,1)+\\
  &&\hskip+5mm\mathrm{Tr}_{2}\,\mathcal{N}_{\mathrm{int}}^{\ast}(1,2)
     \prod_{i_1=1}^{2}\mathcal{G}_{1}^{\ast}(t,i_1)(1+g_{2}(1,2))
     \prod_{i_2=1}^{2}(\mathcal{G}_{1}^{\ast})^{-1}(t,i_2)f_{1}(t,1)f_{1}(t,2),\nonumber\\
     \nonumber\\
\label{Vkeci}
  &&\hskip-5mmf_{1}(t)|_{t=0}=f_{1}^0.
\end{eqnarray}
In expressions (\ref{pc})) and kinetic equation (\ref{Vkec}) it was used: the units where
$h={2\pi\hbar}=1$ is a Planck constant, $m=1$ is the mass of particles and the group
of operators $\mathcal{G}^{\ast}_1(t,j)$ of the free motion of $jth$ particle defined on
the space of trace class operators $f_1$:
\begin{eqnarray}\label{grfm}
    &&\mathbb{R}^1\ni t\mapsto\mathcal{G}^{\ast}_1(t)f_1\doteq e^{-itK(j)}f_1 e^{itK(j)},
\end{eqnarray}
where the operator $K(j)$ is the kinetic energy operator of the $jth$ particle.
The inverse group to the group $(\mathcal{G}_{1}^{\ast})(t)$ we denote by
$(\mathcal{G}_{1}^{\ast})^{-1}(t)=\mathcal{G}_{1}^{\ast}(-t)$. On its domain of
definition the von Neumann operator $\mathcal{N}^{\ast}(1)$ of the free motion is defined
as follows: $\mathcal{N}^{\ast}(1)f_1\doteq-i\,(K(1)f_1-f_1K(1))$, and the operator
$\mathcal{N}^{\ast}_{\mathrm{int}}(1,2)$ is defined by the operator of a two-body
interaction potential $\Phi$, respectively,
$\mathcal{N}^{\ast}_{\mathrm{int}}(1,2)f_2\doteq-i\,(\Phi(1,2)f_2-f_2\Phi(1,2))$.

Thus, it was established that mean field dynamics does not create new correlations except
of those that generating by initial correlations (\ref{pc}).

We remark that the conventional approach to the problem of the description of the propagation
of initial chaos, i.e. in case of initial states specified by a one-particle density operator
without correlation operators, is based on the consideration of an asymptotic behavior of a
solution of the quantum BBGKY hierarchy for marginal density operators constructed within the
framework of the perturbation theory \cite{Go13}--\cite{CG} (for collisional dynamics of hard
spheres see also \cite{CGP}--\cite{G13}).

In this paper we consider the problem of the rigorous description of the evolution of states
of large particle quantum systems within the framework of marginal correlation operators. In
next section 2, we construct a nonperturbative solution of the Cauchy problem of the hierarchy
of nonlinear evolution equations for marginal correlation operators. Then in section 3, we
consider an approach to the description of correlations by means of a one-particle (marginal)
density operator. In section 4, we establish a mean field asymptotic behavior of the
constructed correlation operators. Finally, in section 5, we conclude with some advances of
the developed approaches to the description of quantum correlations.


\section{A nonperturbative solution of the hierarchy of evolution equations for marginal correlation
        operators}

It is known \cite{G12}, that the evolution of states of large particle quantum systems can be
described within the framework of marginal ($s$-particle) density operators as well as in terms
of marginal correlation operators. Traditionally marginal correlation operators are introduced
by means of the cluster expansions of marginal density operators. The physical interpretation
of marginal correlation operators is that the macroscopic characteristics of fluctuations of
mean values of observables are determined by them on the microscopic level. In this section we
construct a solution of the Cauchy problem of the fundamental evolution equations for marginal
correlation operators.

\subsection{Preliminaries: dynamics of correlations}
Let the space $\mathcal{H}$ be a one-particle Hilbert space, then the $n$-particle space
$\mathcal{H}_n=\mathcal{H}^{\otimes n}$ is a tensor product of $n$ Hilbert spaces $\mathcal{H}$.
We adopt the usual convention that $\mathcal{H}^{\otimes 0}=\mathbb{C}$. The Fock space over
the Hilbert space $\mathcal{H}$ we denote by
$\mathcal{F}_{\mathcal{H}}={\bigoplus\limits}_{n=0}^{\infty}\mathcal{H}_{n}$.
A self adjoint operator $f_{n}$ defined on the $n$-particle Hilbert space
$\mathcal{H}_{n}=\mathcal{H}^{\otimes n}$ will be also denoted by the symbol $f_{n}(1,\ldots,n)$.

Let $\mathfrak{L}^{1}(\mathcal{H}_{n})$ be the space of trace class operators
$f_{n}\equiv f_{n}(1,\ldots,n)\in\mathfrak{L}^{1}(\mathcal{H}_{n})$ that satisfy
the symmetry condition: $f_{n}(1,\ldots,n)=f_{n}(i_{1},\ldots,i_{n})$ for arbitrary
$(i_{1},\ldots,i_{n})\in(1,\ldots,n)$, and equipped with the norm:
$\|f_{n}\|_{\mathfrak{L}^{1}(\mathcal{H}_{n})}=\mathrm{Tr}_{1,\ldots,n}|f_{n}(1,\ldots,n)|$,
where $\mathrm{Tr}_{1,\ldots,n}$ are partial traces over $1,\ldots,n$ particles. We denote
by $\mathfrak{L}^{1}_0(\mathcal{H}_{n})$ the everywhere dense set of finite sequences of
degenerate operators with infinitely differentiable kernels with compact supports.

On the space of trace class operators $\mathfrak{L}^1(\mathcal{H}_{n})$ it is defined
the one-parameter mapping $\mathcal{G}^{\ast}_n(t)$
\begin{eqnarray}\label{grG}
    &&\mathbb{R}^1\ni t\mapsto\mathcal{G}^{\ast}_n(t)f_n\doteq e^{-itH_{n}}f_n e^{itH_{n}},
\end{eqnarray}
where the operator $H_{n}$ is the Hamiltonian of a system of $n$ particles, obeying
Maxwell -- Boltzmann statistics, and we use units where $h={2\pi\hbar}=1$ is a Planck
constant and $m=1$ is the mass of particles. The inverse group to the group
$(\mathcal{G}_{n}^{\ast})(t)$ we denote by $(\mathcal{G}_{n}^{\ast})^{-1}(t)=\mathcal{G}_{n}^{\ast}(-t)$.
On its domain of the definition the infinitesimal generator $\mathcal{N}^{\ast}_{n}$ of
the group of operators (\ref{grG}) is determined in the sense of the strong convergence
of the space $\mathfrak{L}^1(\mathcal{H}_{n})$ by the operator
\begin{eqnarray}\label{infOper1}
    &&\lim\limits_{t\rightarrow 0}\frac{1}{t}\big(\mathcal{G}^{\ast}_n(t)f_n-f_n \big)
    =-i\,(H_n f_n - f_n H_n)\doteq\mathcal{N}^{\ast}_n f_n,
\end{eqnarray}
that has the following structure: $\mathcal{N}^{\ast}_n=\sum_{j=1}^{n}\mathcal{N}^{\ast}(j)+
\epsilon\sum_{j_{1}<j_{2}=1}^{n}\mathcal{N}^{\ast}_{\mathrm{int}}(j_{1},j_{2})$, where the
operator $\mathcal{N}^{\ast}(j)$ is a free motion generator of the von Neumann equation
\cite{G12}, the operator $\mathcal{N}^{\ast}_{\mathrm{int}}$ is defined by means of the
operator of a two-body interaction potential $\Phi$ by the formula:
$\mathcal{N}^{\ast}_{\mathrm{int}}(j_{1},j_{2})f_n\doteq -i\,(\Phi(j_{1},j_{2})f_n-f_n \Phi(j_{1},j_{2}))$,
a scaling parameter we denote by $\epsilon>0$.

On the space
$\mathfrak{L}^{1}(\mathcal{F}_\mathcal{H})=\oplus_{n=0}^{\infty}\mathfrak{L}^{1}(\mathcal{H}_{n})$
of sequences $f=(f_0,f_{1},\ldots,f_{n},\ldots)$ of trace class operators
$f_{n}\in\mathfrak{L}^{1}(\mathcal{H}_{n})$ and $f_0\in\mathbb{C}$ the following nonlinear
one-parameter mapping is defined:
\begin{eqnarray}\label{rozvNh}
    &&\mathcal{G}(t;1,\ldots,s\mid f)\doteq\sum\limits_{\mathrm{P}:\,(1,\ldots,s)=\bigcup_j X_j}
      \mathfrak{A}_{|\mathrm{P}|}(t,\{X_1\},\ldots,\{X_{|\mathrm{P}|}\})
      \prod_{X_j\subset \mathrm{P}}f_{|X_j|}(X_j),\quad s\geq1,
\end{eqnarray}
where the symbol $\sum_{\mathrm{P}:\,(1,\ldots,s)=\bigcup_j X_j}$ means the sum over all possible
partitions $\mathrm{P}$ of the set $(1,\ldots,s)$ into $|\mathrm{P}|$ nonempty mutually disjoint
subsets $X_j\subset Y\equiv(1,\ldots,s)$, the set $(\{X_1\},\ldots,\{X_{|\mathrm{P}|}\})$ consists from
elements which are subsets $X_j\subset Y$, i.e. $|(\{X_1\},\ldots,\{X_{|\mathrm{P}|}\})|=|\mathrm{P}|$.
The generating operator $\mathfrak{A}_{|\mathrm{P}|}(t)$ of expansion (\ref{rozvNh}) is the
$|\mathrm{P}|th$-order cumulant of groups of operators (\ref{grG}) defined by the following
expansion:
\begin{eqnarray} \label{cumulantP}
   &&\hskip-8mm \mathfrak{A}_{|\mathrm{P}|}(t,\{X_1\},\ldots,\{X_{|\mathrm{P}|}\})\doteq\\
   &&\sum\limits_{\mathrm{P}^{'}:\,(\{X_1\},\ldots,\{X_{|\mathrm{P}|}\})=
      \bigcup_k Z_k}(-1)^{|\mathrm{P}^{'}|-1}({|\mathrm{P}^{'}|-1})!
      \prod\limits_{Z_k\subset\mathrm{P}^{'}}\mathcal{G}^{\ast}_{|\theta(Z_{k})|}(t,\theta(Z_{k})),\nonumber
\end{eqnarray}
where the declusterization mapping $\theta$ is defined as follows:
$\theta(\{X_1\},\ldots,\{X_{|\mathrm{P}|}\})=(1,\ldots,s)$.

Below we adduce the examples of mapping expansion (\ref{rozvNh}):
\begin{eqnarray*}
    &&\hskip-8mm\mathcal{G}(t;1\mid f)=\mathfrak{A}_{1}(t,1)f_{1}(1),\\
    &&\hskip-8mm\mathcal{G}(t;1,2\mid f)=\mathfrak{A}_{1}(t,\{1,2\})f_{2}(1,2)+
      \mathfrak{A}_{1+1}(t,1,2)f_{1}(1)f_{1}(2),\\
    &&\hskip-8mm\mathcal{G}(t;1,2,3\mid f)=\mathfrak{A}_{1}(t,\{1,2,3\})f_{3}(1,2,3)+
      \mathfrak{A}_{1+1}(t,1,\{2,3\})f_{1}(1)f_{2}(2,3)+\\
    &&\hskip-5mm\mathfrak{A}_{1+1}(t,2,\{1,3\})f_{1}(2)f_{2}(1,3)+
      \mathfrak{A}_{1+1}(t,3,\{1,2\})f_{1}(3)f_{2}(1,2)+
      \mathfrak{A}_{3}(t,1,2,3)f_{1}(1)f_{1}(2)f_{1}(3),
\end{eqnarray*}

On operators $f_{s}\in\mathfrak{L}^{1}(\mathcal{H}_{s}),\,s\geq1$, the mapping $\mathcal{G}(t;Y|f)$
is defined and, according to the inequality for cumulant (\ref{cumulantP}) of groups of operators (\ref{grG})
\begin{eqnarray*}
   &&\Big\|\mathfrak{A}_{|\mathrm{P}|}(t,\{X_1\},\ldots,
      \{X_{|\mathrm{P}|}\})f_s\Big\|_{\mathfrak{L}^{1}(\mathcal{H}_{s})}
      \leq |\mathrm{P}|!\,e^{|\mathrm{P}|}\big\|f_s\big\|_{\mathfrak{L}^{1}(\mathcal{H}_{s})},
\end{eqnarray*}
the following estimate is true:
\begin{eqnarray}\label{gEstimate}
   &&\Big\|\mathcal{G}(t;1,\ldots,s\mid f)\Big\|_{\mathfrak{L}^{1}(\mathcal{H}_{s})}\leq s!e^{2s}c^{s},
\end{eqnarray}
where
$c\equiv e^{3}\max(1,\max_{\mathrm{P}:\,Y=\bigcup_iX_i}\|f_{|X_{i}|}\|_{\mathfrak{L}^{1}(\mathcal{H}_{|X_{i}|})})$.

On the space $\mathfrak{L}^{1}(\mathcal{F}_\mathcal{H})$ one-parameter mapping (\ref{rozvNh})
is a bounded strong continuous group of nonlinear operators and it is determined a solution of the Cauchy
problem for the von Neumann hierarchy for correlation operators \cite{G09}.

The evolution of all possible states of quantum systems of a non-fixed (i.e., arbitrary but finite)
number of identical particles, obeying the Maxwell -- Boltzmann statistics, can be described
by means of the sequence $g(t)=(g_0,g_{1}(t),\ldots,g_{s}(t),\ldots)\in\mathfrak{L}^{1}(\mathcal{F}_\mathcal{H})$
of the correlation operators $g_{s}(t)\equiv g_{s}(t,1,\ldots,s),\,s\geq1$, governed by the Cauchy
problem of the von Neumann hierarchy \cite{GS08}:
\begin{eqnarray}
 \label{vNh}
   &&\frac{\partial}{\partial t}g_{s}(t,1,\ldots,s)=\mathcal{N}(1,\ldots,s\mid g(t)),\\
   \nonumber\\
 \label{vNhi}
   &&g_{s}(t)\big|_{t=0}=g_{s}^{0,\epsilon},\quad s\geq1,
\end{eqnarray}
where $\epsilon>0$ is a scaling parameter. The generator of the hierarchy of nonlinear
evolution equations (\ref{vNh}) has the structure
\begin{eqnarray}\label{vNgenerator}
   &&\hskip-5mm \mathcal{N}(1,\ldots,s\mid g(t))\doteq\mathcal{N}^{\ast}_{s}g_{s}(t,1,\ldots,s)+\\
   &&\hskip+5mm \epsilon\sum\limits_{\mathrm{P}:\,(1,\ldots,s)=X_{1}\bigcup X_2}\,\sum\limits_{i_{1}\in X_{1}}
      \sum\limits_{i_{2}\in X_{2}}\mathcal{N}_{\mathrm{int}}^{\ast}(i_{1},i_{2})
      g_{|X_{1}|}(t,X_{1})g_{|X_{2}|}(t,X_{2}),\nonumber
\end{eqnarray}
where the symbol ${\sum\limits}_{\mathrm{P}:\,(1,\ldots,s)=X_{1}\bigcup X_2}$ means the sum over all
possible partitions $\mathrm{P}$ of the set $Y\equiv(1,\ldots,s)$ into two nonempty mutually disjoint
subsets $X_1\subset Y$ and $X_2\subset Y$, and the operator $\mathcal{N}^{\ast}_{s}$ is the von Neumann
operator defined by formula (\ref{infOper1}) on the subspace
$\mathfrak{L}^{1}_0(\mathcal{H}_s)\subset\mathfrak{L}^{1}_0(\mathcal{H}_s)$.

We remark that correlation operators are introduced by means of the cluster expansions of the density
operators (the kernel of a density operator is known as a density matrix) governed by the von Neumann
equations, and it is to enable to describe of the evolution of states by the equivalent method
in comparison with the density operators \cite{G12}.

A nonperturbative solution of the Cauchy problem of the von Neumann hierarchy (\ref{vNh}),(\ref{vNhi})
for correlation operators is determined by the group of nonlinear operators (\ref{rozvNh}), i.e.
\begin{eqnarray}\label{ghs}
   &&g(t,1,\ldots,s)=\mathcal{G}(t;1,\ldots,s\mid g(0)), \quad s\geq1,
\end{eqnarray}
where $g(0)=(g_0,g_{1}^{0,\epsilon},\ldots,g_{n}^{0,\epsilon},\ldots)$ is a sequence of initial
correlation operators.

We note that in case of the absence of correlations between particles at initial time, i.e.
initial data (\ref{vNhi}), satisfying a chaos condition, the sequence of initial correlation
operators has the form
\begin{eqnarray*}\label{gChaos}
   &&g(0)=(0,g_{1}^{0,\epsilon},0,\ldots,0,\ldots).
\end{eqnarray*}
Then solution (\ref{ghs}) of the Cauchy problem of the von Neumann hierarchy (\ref{vNh}),(\ref{vNhi})
is represented by the following expansions:
\begin{eqnarray*}\label{gth}
   &&g_{s}(t,1,\ldots,s)=\mathfrak{A}_{s}(t,1,\ldots,s)\,\prod\limits_{i=1}^{s}g_{1}^{0,\epsilon}(i),\quad s\geq1,
\end{eqnarray*}
where the operator $\mathfrak{A}_{s}(t)$ is the $sth$-order cumulant of groups of operators (\ref{grG})
determined by the expansion
\begin{eqnarray}\label{cumcp}
   &&\mathfrak{A}_{s}(t,1,\ldots,s)=
      \sum\limits_{\mathrm{P}:\,(1,\ldots,s)=\bigcup_i X_i}(-1)^{|\mathrm{P}|-1}({|\mathrm{P}|-1})!
      \prod\limits_{X_i\subset\mathrm{P}}\mathcal{G}^{\ast}_{|X_i|}(t,X_i),
\end{eqnarray}
and we used notations accepted in formula (\ref{rozvNh}).

\subsection{A nonperturbative solution of the nonlinear BBGKY hierarchy for marginal correlation
         operators}
We introduce the marginal correlation operators by means of the macroscopic characteristics of fluctuations
of mean values of observables are determined.

The marginal correlation operators are defined within the framework of a solution of the Cauchy
problem of the von Neumann hierarchy (\ref{vNh}),(\ref{vNhi}) by the following series expansions:
\begin{eqnarray}\label{Gexpsg}
   &&G_{s}(t,1,\ldots,s)\doteq\sum\limits_{n=0}^{\infty}\frac{1}{n!}\,
      \mathrm{Tr}_{s+1,\ldots,s+n}\,\,\mathcal{G}(t;1,\ldots,s+n\mid g(0)),\quad s\geq1.
\end{eqnarray}
According to estimate (\ref{gEstimate}), series (\ref{Gexpsg}) exists and the following estimate holds:
$\big\|G_s(t)\big\|_{\mathfrak{L}^{1}(\mathcal{H}_{s})}\leq s!(2e^2)^s\emph{c}^s\sum_{n=0}^{\infty}(2e^2)^n\emph{c}^n$.

The evolution of all possible states of quantum large particle systems, obeying the Maxwell -- Boltzmann
statistics, can be described by means of the sequence
$G(t)=(I,G_1(t),G_2(t),\ldots,$  $G_s(t),\ldots)\in\mathfrak{L}^{1}(\mathcal{F}_\mathcal{H})$
of marginal correlation operators governed by the Cauchy problem of the following hierarchy
of nonlinear evolution equations (the nonlinear quantum BBGKY hierarchy):
\begin{eqnarray}
 \label{gBigfromDFBa}
   &&\hskip-5mm\frac{\partial}{\partial t}G_s(t,Y)=\mathcal{N}(Y\mid G(t))+
      \mathrm{Tr}_{s+1}\sum_{i\in Y}\mathcal{N}^{\ast}_{\mathrm{int}}(i,s+1)\big(G_{s+1}(t,Y,s+1)+\\
   &&\hskip+5mm\epsilon\sum_{\mbox{\scriptsize$\begin{array}{c}\mathrm{P}:(Y,s+1)=X_1\bigcup X_2,\\i\in
      X_1;s+1\in X_2\end{array}$}}G_{|X_1|}(t,X_1)G_{|X_2|}(t,X_2)\big),\nonumber\\ \nonumber\\
 \label{gBigfromDFBai}
   &&\hskip-5mmG_{s}(t)\big|_{t=0}=G_{s}^{0,\epsilon}, \quad s\geq1.
\end{eqnarray}
In evolution equations (\ref{gBigfromDFBa}) it was used notations accepted above, in particular,
$(1,\ldots,s)\equiv Y$, the operators $\mathcal{N}(1,\ldots,s\mid G(t))\equiv\mathcal{N}(Y\mid G(t)),\,s\geq1,$
are generators (\ref{vNgenerator}) of the von Neumann hierarchy (\ref{vNh}) and $\epsilon>0$ is a scaling parameter.

The rigorous derivation of the hierarchy of evolution equations for marginal correlation
operators (\ref{gBigfromDFBa}), according to definition (\ref{Gexpsg}), consists in its derivation
from the von Neumann hierarchy for correlation operators (\ref{vNh}) \cite{GP13}
(for marginal density operators see paper \cite{P10}).

A nonperturbative solution of the Cauchy problem (\ref{gBigfromDFBa}),(\ref{gBigfromDFBai})
is represented by a sequence of the following self-adjoint operators:
\begin{eqnarray}\label{sss}
    &&G_{s}(t,Y)=\sum\limits_{n=0}^{\infty}\frac{1}{n!}
        \,\mathrm{Tr}_{s+1,\ldots,s+n}\,\mathfrak{A}_{1+n}(t;\{Y\},s+1,\ldots,s+n\mid G(0)),\quad s\geq1,
\end{eqnarray}
where a sequence of initial marginal correlation operators we denote by
$G(0)=(I,G_1^{0,\epsilon}(1),\ldots,$ $G_s^{0,\epsilon}(1,\ldots,s),\ldots)$. The generating operators of
series expansion (\ref{sss}), in particular, the operator $\mathfrak{A}_{1+n}(t;\{Y\},s+1,\ldots,s+n\mid G(0))$
is the $(1+n)th$-order cumulant of groups of nonlinear operators (\ref{rozvNh}) of the von Neumann
hierarchy for correlation operators
\begin{eqnarray} \label{cc}
   &&\hskip-8mm \mathfrak{A}_{1+n}(t;\{Y\},s+1,\ldots,s+n\mid G(0))\doteq\\
   &&\sum\limits_{\mathrm{P}:\,(\{Y\},s+1,\ldots,s+n)=
      \bigcup_k X_k}(-1)^{|\mathrm{P}|-1}({|\mathrm{P}|-1})!\,
      \mathcal{G}(t;\theta(X_1)\mid\ldots\mathcal{G}(t;\theta(X_{|\mathrm{P}|})\mid G(0))\ldots),\nonumber
\end{eqnarray}
where the composition of mappings (\ref{rozvNh}) of corresponding noninteracting groups of particles
is denoted by $\mathcal{G}(t;\theta(X_1)\mid \ldots\mathcal{G}(t;\theta(X_{|\mathrm{P}|})\mid G(0))\ldots)$, for example,
\begin{eqnarray*}
    &&\hskip-5mm \mathcal{G}\big(t;1\mid\mathcal{G}(t;2\mid f)\big)=\mathfrak{A}_{1}(t,1)\mathfrak{A}_{1}(t,2)f_{2}(1,2),\\
    &&\hskip-5mm \mathcal{G}\big(t;1,2\mid\mathcal{G}(t;3\mid f)\big)=
        \mathfrak{A}_{1}(t,\{1,2\})\mathfrak{A}_{1}(t,3)f_{3}(1,2,3)+\\
    &&\hskip+15mm\mathfrak{A}_{2}(t,1,2)\mathfrak{A}_{1}(t,3)\big(f_{1}(1)f_{2}(2,3)+f_{1}(2)f_{2}(1,3)\big).
\end{eqnarray*}

Below we adduce the examples of expansions (\ref{cc}). The first order cumulant of the groups
of nonlinear operators (\ref{rozvNh}) is the same group of nonlinear operators, i.e.
\begin{eqnarray*}
     &&\mathfrak{A}_{1}(t;\{1,\ldots,s\}\mid G(0))=\mathcal{G}(t;1,\ldots,s \mid G(0)),
\end{eqnarray*}
in case of $s=2$ the second order cumulant of groups of nonlinear operators (\ref{rozvNh})
represents by the following expansion:
\begin{eqnarray*}
     &&\hskip-8mm \mathfrak{A}_{1+1}(t;\{1,2\},3\mid G(0))=\mathcal{G}(t;1,2,3\mid G(0))-
       \mathcal{G}\big(t;1,2\mid\mathcal{G}(t;3\mid G(0))\big)=\\
     &&\mathfrak{A}_{1+1}(t,\{1,2\},3)G^{0,\epsilon}_{3}(1,2,3)+\big(\mathfrak{A}_{1+1}(t,\{1,2\},3)-
        \mathfrak{A}_{1+1}(t,2,3)\mathfrak{A}_{1}(t,1)\big)G^{0,\epsilon}_{1}(1)G^{0,\epsilon}_{2}(2,3)+\\
     &&\big(\mathfrak{A}_{1+1}(t,\{1,2\},3)-
        \mathfrak{A}_{1+1}(t,1,3)\mathfrak{A}_{1}(t,2)\big)G^{0,\epsilon}_{1}(2)G^{0,\epsilon}_{2}(1,3)+\\
     &&\mathfrak{A}_{1+1}(t,\{1,2\},3)G^{0,\epsilon}_{1}(3)G^{0,\epsilon}_{2}(1,2)+
        \mathfrak{A}_{3}(t,1,2,3)G^{0,\epsilon}_{1}(1)G^{0,\epsilon}_{1}(2)G^{0,\epsilon}_{1}(3),
\end{eqnarray*}
where the operator $\mathfrak{A}_{3}(t,1,2,3)=\mathfrak{A}_{1+1}(t,\{1,2\},3)-\mathfrak{A}_{1+1}(t,2,3)\mathfrak{A}_{1}(t,1)-
\mathfrak{A}_{1+1}(t,1,3)\mathfrak{A}_{1}(t,2)$ is the third order cumulant of groups of operators (\ref{grG}).

In case of initial data specified by the sequence of marginal correlation operators
\begin{eqnarray}\label{inscc}
   &&\hskip-8mm G^{(c)}=\big(0,G_1^{0,\epsilon},0,\ldots,0,\ldots\big),
\end{eqnarray}
i.e. initial data satisfying a chaos condition, according to definition (\ref{cc}),
marginal correlation operators (\ref{sss}) are represented by the following series
expansions:
\begin{eqnarray}\label{mcc}
   &&G_{s}(t,1,\ldots,s)=\sum\limits_{n=0}^{\infty}\frac{1}{n!}
       \,\mathrm{Tr}_{s+1,\ldots,s+n}\,\mathfrak{A}_{s+n}(t;1,\ldots,s+n)
       \prod_{i=1}^{s+n}G_1^{0,\epsilon}(i),\quad s\geq1,
\end{eqnarray}
where the generating operator $\mathfrak{A}_{s+n}(t)$ of this series is the $(s+n)th$-order
cumulant (\ref{cumcp}) of groups of operators (\ref{grG}).

We remark that within the framework of marginal density operators defined by means of the cluster
expansions of marginal correlation operators
\begin{eqnarray*}
   &&F_{s}^{0,\epsilon}(1,\ldots,s)=
      \sum_{\mbox{\scriptsize $\begin{array}{c}\mathrm{P}:(1,\ldots,s)=\bigcup_{i}X_{i}\end{array}$}}
      {\prod\limits}_{X_i\subset\mathrm{P}}G_{|X_i|}^{0,\epsilon}(X_i),\quad s\geq1,
\end{eqnarray*}
initial state similar to (\ref{inscc}) is specified by the sequence
$F^{(c)}=\big(I,F_1^{0,\epsilon}(1),\ldots,{\prod\limits}_{i=1}^{n}F_1^{0,\epsilon}(i),\ldots\big)$,
and in case of sequence (\ref{mcc}) the marginal density operators are represented by the following
series expansions (a nonperturbative solution of the quantum BBGKY hierarchy \cite{G12}):
\begin{eqnarray*}
   &&F_{s}(t,1,\ldots,s)=\sum\limits_{n=0}^{\infty}\frac{1}{n!}
       \,\mathrm{Tr}_{s+1,\ldots,s+n}\,\mathfrak{A}_{1+n}(t;\{Y\},s+1,\ldots,s+n)
       \prod_{i=1}^{s+n}F_1^{0,\epsilon}(i),\quad s\geq1,
\end{eqnarray*}
where the generating operator $\mathfrak{A}_{1+n}(t)$ is the $(1+n)th$-order cumulant of groups
of operators (\ref{grG}).

One of the methods to derive the series expansion (\ref{sss}) for marginal correlation operators
consists on the application of the cluster expansions of groups of nonlinear operators (\ref{rozvNh})
over cumulants (\ref{cc}) in the definition of marginal correlation operators (\ref{Gexpsg}) and the
sequence of initial correlation operators $g(0)=(I,g_1^{0,\epsilon}(1),\ldots,g_n^{0,\epsilon}(1,\ldots,n),\ldots)$
determined by means of the marginal correlation operators:
\begin{eqnarray}\label{G0}
   &&g_{s}^{0,\epsilon}(1,\ldots,s)\doteq\sum\limits_{n=0}^{\infty}(-1)^n\frac{1}{n!}\,
      \mathrm{Tr}_{s+1,\ldots,s+n}\,\,G_{s+n}^{0,\epsilon}(1,\ldots,s+n),\quad s\geq1.
\end{eqnarray}
Indeed, developing the generating operators of series (\ref{sss}) as cluster expansions
\begin{eqnarray}\label{RR}
   &&\hskip-8mm \mathcal{G}(t;1,\ldots,s+n\mid f)=\sum\limits_{\mathrm{P}:\,(1,\ldots,s+n)=\bigcup_k X_k}
      \mathfrak{A}_{|X_1|}(t;X_1\mid\ldots\mathfrak{A}_{|X_{|\mathrm{P}|}|}(t;X_{|\mathrm{P}|}\mid f)\ldots),
\end{eqnarray}
according to definition (\ref{G0}), we derive expressions (\ref{sss}). The solutions of recursive
relations (\ref{RR}) is represented by expansions (\ref{cc}).

We remark that on the space $\mathfrak{L}^{1}(\mathcal{F}_\mathcal{H})$ the generating operator
(\ref{cc}) of series expansion (\ref{sss}) can be represented as the $(1+n)th$-order reduced
cumulant of groups of nonlinear operators (\ref{rozvNh}) of the von Neumann hierarchy \cite{GP13}
\begin{eqnarray}\label{ssss}
    &&\hskip-8mm U_{1+n}(t;\{1,\ldots,s\},s+1,\ldots,s+n \mid G(0))\doteq\\
    &&\sum\limits_{k=0}^{n}(-1)^k \frac{n!}{k!(n-k)!}\,\sum\limits_{\mathrm{P}:\,
       (\theta(\{1,\ldots,s\}),s+1,\ldots,s+n-k)=\bigcup_i X_i}
       \mathfrak{A}_{|\mathrm{P}|}\big(t,\{X_1\},\ldots,\{X_{|\mathrm{P}|}\}\big)\times\nonumber\\
    &&\sum\limits_{k_1=0}^{k}\frac{k!}{k_{1}!(k-k_{1})!}\ldots
       \sum\limits_{k_{|\mathrm{P}|-1}=0}^{k_{|\mathrm{P}|-2}}
       \frac{k_{|\mathrm{P}|-2}!}{k_{|\mathrm{P}|-1}!(k_{|\mathrm{P}|-2}-k_{|\mathrm{P}|-1})!}
       G_{|X_1|+k-k_1}^{0,\epsilon}(X_1,\nonumber\\
    && s+n-k+1,\ldots,s+n-k_1)\ldots G_{|X_{|\mathrm{P}|}|+k_{|\mathrm{P}|-1}}^{0,\epsilon}(X_{|\mathrm{P}|},
       s+n-k_{|\mathrm{P}|-1}+1,\ldots,s+n).\nonumber
\end{eqnarray}
We adduce simplest examples of reduced cumulants (\ref{ssss}) of groups of nonlinear operators (\ref{rozvNh}):
\begin{eqnarray*}
    &&\hskip-8mm U_{1}(t;\{1,\ldots,s\}\mid G(0))=\mathcal{G}(t;1,\ldots,s\mid G(0))=\\
    &&\sum\limits_{\mathrm{P}:\,(1,\ldots,s)=
        \bigcup_i X_i}\mathfrak{A}_{|\mathrm{P}|}\big(t,\{X_1\},\ldots,\{X_{|\mathrm{P}|}\}\big)
        \prod\limits_{X_i\subset\mathrm{P}}G_{|X_i|}^{0,\epsilon}(X_{i}),\\
    &&\hskip-8mm U_{1+1}(t;\{1,\ldots,s\},s+1 \mid G(0))=\sum\limits_{\mathrm{P}:\,(1,\ldots,s+1)=\bigcup_i X_i}
        \mathfrak{A}_{|\mathrm{P}|}\big(t,\{X_1\},\ldots,\{X_{|\mathrm{P}|}\}\big)
        \prod\limits_{X_i\subset\mathrm{P}}G_{|X_i|}^{0,\epsilon}(X_{i})-\\
    &&\sum\limits_{\mathrm{P}:\,(1,\ldots,s)=\bigcup_i X_i}
        \mathfrak{A}_{|\mathrm{P}|}\big(t,\{X_1\},\ldots,\{X_{|\mathrm{P}|}\}\big)\sum_{j=1}^{|\mathrm{P}|}
        G_{|X_j|+1}^{0,\epsilon}(X_{j},s+1)
        \prod\limits_{\mbox{\scriptsize $\begin{array}{c}{X_i\subset\mathrm{P}},\\X_i\neq X_j\end{array}$}}
        G_{|X_i|}^{0,\epsilon}(X_{i}).
\end{eqnarray*}

We note also that nonperturbative solution of the nonlinear quantum BBGKY hierarchy (\ref{sss}) or in
the form of series expansions (\ref{ssss}) can be transformed to the perturbation (iteration) series
as a result of the application of analogs of the Duhamel equation to cumulants (\ref{cumulantP}) of
groups of operators (\ref{grG}).

The following statement is true.
\begin{theorem}
If $\max_{n\geq1}\big\|G_n^{0,\epsilon}\big\|_{\mathfrak{L}^{1}(\mathcal{H}_{n})}<(2e^{3})^{-1}$,
then in case of bounded interaction potentials for $t\in\mathbb{R}$ a solution of the Cauchy problem
of the nonlinear quantum BBGKY hierarchy (\ref{gBigfromDFBa}),(\ref{gBigfromDFBai}) is determined
by a sequence of marginal correlation operators represented by series expansions (\ref{sss}). If $G_{n}^{0,\epsilon}\in\mathfrak{L}^{1}_{0}(\mathcal{H}_{n})\subset\mathfrak{L}^{1}(\mathcal{H}_{n})$,
it is a strong solution and for arbitrary initial data
$G_{n}^{0,\epsilon}\in\mathfrak{L}^{1}(\mathcal{H}_{n})$ it is a weak solution.
\end{theorem}
The proof of the existence theorem is similar to the case of the reduced representation of a
nonperturbative solution of the nonlinear quantum BBGKY hierarchy \cite{GP13}.


\section{On the representation of marginal correlation operators by means of a one-particle density operator}

In case of initial states specified in terms of a one-particle (marginal) density
operator and correlation operators the evolution of all possible states of quantum
large particle systems can be described within the framework of a one-particle
density operator governed by the kinetic equation without any approximations. In
this section we consider an approach to the description of the processes of the
creation correlations and the propagation of initial correlations by means of a
one-particle density operator that is a solution of the generalized quantum kinetic
equation with initial correlations.

\subsection{Marginal correlation functionals}
If initial states specified in terms of a one-particle density operator (\ref{inscc}),
then the evolution of states given in the framework of the sequence $G(t)=(I,G_1(t),\ldots,G_s(t),\ldots)$
of marginal correlation operators (\ref{sss}) can be described by the sequence
$G(t\mid G_{1}(t))=(I,G_1(t),G_2(t\mid G_{1}(t)),$ $\ldots,G_s(t\mid G_{1}(t)),\ldots)$
of marginal correlation functionals $G_s(t,1,\ldots,s\mid G_{1}(t)),\,s\geq2$, with respect
to the one-particle density (correlation) operator $G_1(t)$ \cite{GT10}.

In case of initial states (\ref{inscc}) the marginal correlation functionals
$G_{s}\big(t\mid G_{1}(t)\big),\,s\geq2$, are represented by the series expansions:
\begin{eqnarray}\label{cf}
    &&\hskip-15mm G_{s}(t,1,\ldots,s\mid G_{1}(t))=
        \sum\limits_{n=0}^{\infty}\frac{1}{n!}\,\mathrm{Tr}_{s+1,\ldots,s+n}
        \mathfrak{V}_{s+n}(t,\theta(\{Y\}),s+1,\ldots,s+n)\prod _{i=1}^{s+n}G_{1}(t,i),
\end{eqnarray}
In this formula it is used the notion of the declusterization mapping: $\theta(\{Y\})=Y\equiv(1,\ldots,s)$,
and the one-particle (marginal) correlation operator is determined by the series expansion
\begin{eqnarray*}\label{ske}
   &&\hskip-15mm G_{1}(t,1)=\sum\limits_{n=0}^{\infty}\frac{1}{n!}\,\mathrm{Tr}_{2,\ldots,{1+n}}\,\,
      \mathfrak{A}_{1+n}(t)\prod_{i=1}^{n+1}G_{1}^{0,\epsilon}(i),
\end{eqnarray*}
where the generating operator $\mathfrak{A}_{1+n}(t)\equiv\mathfrak{A}_{1+n}(t,1,\ldots,n+1)$
is the $(1+n)th$-order cumulant (\ref{cumcp}) of groups of operators (\ref{grG}).
We adduce simplest examples of the generating operators of series expansion (\ref{cf}):
\begin{eqnarray*}
   &&\mathfrak{V}_{s}(t,\theta(\{Y\}))=\widehat{\mathfrak{A}}_{s}(t,\theta(\{Y\}))\doteq
       \mathfrak{A}_{s}(t,\theta(\{Y\}))\prod_{i=1}^{s}\mathfrak{A}_{1}^{-1}(t,i),\\
   &&\mathfrak{V}_{s+1}(t,\theta(\{Y\}),s+1)=\widehat{\mathfrak{A}}_{s+1}(t,\theta(\{Y\}),s+1)-
      \widehat{\mathfrak{A}}_{s}(t,\theta(\{Y\}))\sum_{i=1}^s\widehat{\mathfrak{A}}_{2}(t,i,s+1),
\end{eqnarray*}
where the operator $\mathfrak{A}_{1}^{-1}(t)$ is inverse to the operator $\mathfrak{A}_{1}(t)$,
and in particular case $s=2$ we have
\begin{eqnarray*}
    &&\mathfrak{V}_{2}(t,\theta(\{1,2\}))=\widehat{\mathcal{G}}_{2}(t,1,2)-I,
\end{eqnarray*}
where $\widehat{\mathcal{G}}_{2}(t,1,2)\doteq
\mathcal{G}^{\ast}_{2}(t,1,2)(\mathcal{G}^{\ast}_{1})^{-1}(t,1)(\mathcal{G}^{\ast}_{1})^{-1}(t,2)$
is the scattering operator.

A method of the construction of marginal correlation functionals (\ref{cf}) is based on the application
of kinetic cluster expansions \cite{G12} to the generating operators of series (\ref{mcc}). The structure
of generating operators of series (\ref{cf}) we shall define below for more general case of initial states.

We conclude only that the generating operator of the $nth$ term of series expansion (\ref{cf}) of marginal
correlation functionals is the $(s+n)th$-order evolution operator of cumulants of scattering operators.

It should be noted that marginal correlation functionals (\ref{cf}) describe the correlations created
by dynamics of large particle quantum systems by means of a one-particle correlation (density) operator.

Now we consider the case of initial states specified by a one-particle marginal density operator
with correlations, namely, initial states specified by the following sequence of marginal correlation
operators:
\begin{eqnarray}\label{insc}
   &&\hskip-8mm G^{(cc)}=\big(I,G_1^{0,\epsilon}(1),g_{2}^{\epsilon}(1,2)\prod_{i=1}^{2}G_1^{0,\epsilon}(i),
        \ldots,g_{n}^{\epsilon}(1,\ldots,n)\prod_{i=1}^{n}G_1^{0,\epsilon}(i),\ldots\big),
\end{eqnarray}
where the operators
$g_{n}^{\epsilon}(1,\ldots,n)\equiv g_{n}^{\epsilon}\in\mathfrak{L}^{1}_0(\mathcal{H}_n),\,n\geq2$,
are specified the initial correlations. We remark that such assumption about initial states is
intrinsic for the kinetic description of many-particle systems. On the other hand, initial data
(\ref{insc}) is typical for the condensed states of large particle quantum systems, for example,
the equilibrium state of the Bose condensate satisfies the weakening of correlation condition
with the correlations which characterize the condensed state \cite{BQ}.

In this case the marginal correlation functionals $G_s(t\mid G_{1}(t)),\,s\geq2$, are defined with
respect to the one-particle (marginal) correlation operator
\begin{eqnarray}\label{ske}
   &&\hskip-12mm G_{1}(t,1)= \sum\limits_{n=0}^{\infty}\frac{1}{n!}\,\mathrm{Tr}_{2,\ldots,{1+n}}\,\,
      \mathfrak{A}_{1+n}(t)g_{n+1}^{\epsilon}(1,\ldots,n+1)\prod_{i=1}^{n+1}G_{1}^{0,\epsilon}(i),
\end{eqnarray}
where the generating operator $\mathfrak{A}_{1+n}(t)\equiv\mathfrak{A}_{1+n}(t,1,\ldots,n+1)$
is the $(1+n)th$-order cumulant (\ref{cumcp}) of groups of operators (\ref{grG}),
and they are represented by the following series expansions:
\begin{eqnarray}\label{f}
     &&\hskip-12mm G_{s}(t,Y\mid G_{1}(t))\doteq
        \sum _{n=0}^{\infty }\frac{1}{n!}\,\mathrm{Tr}_{s+1,\ldots,{s+n}}\,
        \mathfrak{G}_{s+n}(t,\theta(\{Y\}),X\setminus Y)\prod_{i=1}^{s+n}G_{1}(t,i),\quad s\geq2,
\end{eqnarray}
where the $(s+n)th$-order generating operator $\mathfrak{G}_{s+n}(t),\,n\geq0$, of this series
is determined by the following expansion:
\begin{eqnarray}\label{skrrc}
   &&\hskip-8mm\mathfrak{G}_{s+n}(t,\theta(\{Y\}),X\setminus Y)\doteq
       n!\,\sum_{k=0}^{n}\,(-1)^k\,\sum_{n_1=1}^{n}\ldots
       \sum_{n_k=1}^{n-n_1-\ldots-n_{k-1}}\frac{1}{(n-n_1-\ldots-n_k)!}\times\\
   &&\breve{\mathfrak{A}}_{s+n-n_1-\ldots-n_k}(t,\theta(\{Y\}),s+1,\ldots,
       s+n-n_1-\ldots-n_k)\times\nonumber\\
   &&\prod_{j=1}^k\,\sum\limits_{\mbox{\scriptsize$\begin{array}{c}
       \mathrm{D}_{j}:Z_j=\bigcup_{l_j}X_{l_j},\\
       |\mathrm{D}_{j}|\leq s+n-n_1-\dots-n_j\end{array}$}}\frac{1}{|\mathrm{D}_{j}|!}
       \sum_{i_1\neq\ldots\neq i_{|\mathrm{D}_{j}|}=1}^{s+n-n_1-\ldots-n_j}\,\,
       \prod_{X_{l_j}\subset \mathrm{D}_{j}}\,\frac{1}{|X_{l_j}|!}\,\,
       \breve{\mathfrak{A}}_{1+|X_{l_j}|}(t,i_{l_j},X_{l_j}).\nonumber
\end{eqnarray}
In formula (\ref{skrrc}) we denote: $Y\equiv (1,\ldots,s),\,X\setminus Y\equiv (s+1,\ldots,s+n)$,
the sum over all possible dissections of the linearly ordered set
$Z_j\equiv(s+n-n_1-\ldots-n_j+1,\ldots,s+n-n_1-\ldots-n_{j-1})$ on no more than $s+n-n_1-\ldots-n_j$
linearly ordered subsets we denote by $\sum_{\mathrm{D}_{j}:Z_j=\bigcup_{l_j} X_{l_j}}$ and the
$(s+n)th$-order scattering cumulant is defined by the formula
\begin{eqnarray*}
   &&\breve{\mathfrak{A}}_{s+n}(t,\theta(\{Y\}),X\setminus Y)\doteq
       \mathfrak{A}_{s+n}(t,\theta(\{Y\}),X\setminus Y)g_{s+n}^{\epsilon}(\theta(\{Y\}),X\setminus Y)
       \prod_{i=1}^{s+n}\mathfrak{A}_{1}^{-1}(t,i),
\end{eqnarray*}
where the operator $g_{s+n}^{\epsilon}(\theta(\{Y\}),X\setminus Y)$ is specified initial correlations
(\ref{insc}), and notations accepted above were used. We give examples of the scattering cumulants
\begin{eqnarray*}
   &&\hskip-5mm\mathfrak{G}_{s}(t,\theta(\{Y\}))=\breve{\mathfrak{A}}_{s}(t,\theta(\{Y\}))\doteq
      \mathfrak{A}_{s}(t,\theta(\{Y\}))g_{s}^{\epsilon}(\theta(\{Y\}))
      \prod_{i=1}^{s}\mathfrak{A}_{1}^{-1}(t,i),\\
   &&\hskip-5mm\mathfrak{G}_{s+1}(t,\theta(\{Y\}),s+1)=\mathfrak{A}_{s+1}(t,\theta(\{Y\}),s+1)
      g_{s+1}^{\epsilon}(\theta(\{Y\}),s+1)\prod_{i=1}^{s+1}\mathfrak{A}_{1}^{-1}(t,i)-\\
   &&\mathfrak{A}_{s}(t,\theta(\{Y\}))g_{s}^{\epsilon}(\theta(\{Y\}))
      \prod_{i=1}^{s}\mathfrak{A}_{1}^{-1}(t,i)
      \sum_{j=1}^s\mathfrak{A}_{2}(t,j,s+1)g_{2}^{\epsilon}(j,s+1)\mathfrak{A}_{1}^{-1}(t,j)
      \mathfrak{A}_{1}^{-1}(t,s+1).
\end{eqnarray*}

If $\|G_{1}(t)\|_{\mathfrak{L}^{1}(\mathcal{H})}<e^{-(3s+2)}$, then for arbitrary $t\in \mathbb{R}$
series expansion (\ref{f}) converges in the norm of the space $\mathfrak{L}^{1}(\mathcal{H}_{s})$ \cite{G12}.

We emphasize that marginal correlation functionals (\ref{f}) describe the processes of the creation and the
propagation of correlations generated by dynamics of large particle quantum systems in the presence of initial
correlations by means of a one-particle density operator.

\subsection{A quantum kinetic equation with initial correlations}
We establish the evolution equation for one-particle marginal correlation operator (\ref{ske}).

As a result of the differentiation over the time variable of the operator represented by series
(\ref{ske}) in the sense of the norm convergence of the space $\mathfrak{L}^{1}(\mathcal{H})$,
then due to the application of the kinetic cluster expansions \cite{GTsm} to the generating
operators of obtained series expansion, for one-particle (marginal) correlation operator (\ref{ske})
we derive the following identity
\begin{eqnarray}\label{gkec}
   &&\hskip-9mm\frac{\partial}{\partial t}G_{1}(t,1)= \mathcal{N}^{\ast}(1)G_{1}(t,1)+
      \epsilon\,\mathrm{Tr}_{2}\,\mathcal{N}_{\mathrm{int}}^{\ast}(1,2)G_{1}(t,1)G_{1}(t,2)+\\
   &&\hskip+5mm\epsilon\,\mathrm{Tr}_{2}\,\mathcal{N}_{\mathrm{int}}^{\ast}(1,2)G_{2}(t,1,2\mid G_{1}(t)),\nonumber
\end{eqnarray}
where the second part of the collision integral in equation (\ref{gkec}) is determined in terms of
the marginal correlation functional represented by series expansion (\ref{f}) in case of $s=2$. This
identity we treat as the quantum kinetic equation and we refer to this evolution equation as the
generalized quantum kinetic equation with initial correlations.

We emphasize that the coefficients in an expansion of the collision integral of the non-Markovian
kinetic equation (\ref{gkec}) are determined by the operators specified initial correlations (\ref{insc}).

For the generalized quantum kinetic equation with initial correlations (\ref{gkec})
on the space $\mathfrak{L}^{1}(\mathcal{H})$ the following statement is true.
\begin{theorem}
If $\|G_1^{0,\epsilon}\|_{\mathfrak{L}^{1}(\mathcal{H})}<(e(1+e^{9}))^{-1}$, the global
in time solution of initial-value problem of kinetic equation (\ref{gkec}) is
determined by series expansion (\ref{ske}). For initial data
$G_1^{0,\epsilon}\in\mathfrak{L}^{1}_{0}(\mathcal{H})$ it is a strong solution
and for an arbitrary initial data it is a weak solution.
\end{theorem}
The proof of the existence theorem is similar to the case of the generalized quantum
kinetic equation \cite{GT10}.


\section{A mean field asymptotic behavior of marginal correlation operators}

This section deal with the scaling asymptotic behavior of the constructed marginal correlation operators.
The processes of the creation and the propagation of correlations will be described in a mean field limit
for two cases: initial states satisfying a chaos property and initial states specified by means of a
one-particle density operator and correlations.

\subsection{On the propagation of initial chaos}
In the beginning we give comments on the mean field asymptotic behavior of constructed solution (\ref{sss})
in case of the initial state (\ref{inscc}), satisfying a chaos condition \cite{G12}.

We assume the existence of a mean field limit of the initial marginal correlation operator (or the one-particle
density operator) in the following sense
\begin{eqnarray}\label{asic1}
   &&\lim\limits_{\epsilon\rightarrow 0}\big\|\epsilon G_{1}^{0,\epsilon}-
      g_{1}^{0}\big\|_{\mathfrak{L}^{1}(\mathcal{H})}=0.
\end{eqnarray}

Since $nth$ term of series expansion (\ref{mcc}) for $s$-particle marginal correlation operators
is determined by the $(s+n)th$-order cumulants of asymptotically perturbed groups of operators
(\ref{grG}) for which the following equality is true
\begin{eqnarray}\label{apg}
   &&\lim\limits_{\epsilon\rightarrow0}\Big\|\frac{1}{\epsilon^{n}}\,
     \mathfrak{A}_{s+n}(t,1,\ldots,s+n)f_{s+n}\Big\|_{\mathfrak{L}^{1}(\mathcal{H}_{s+n})}=0,\quad s\geq2,
\end{eqnarray}
then we establish the property of the propagation of initial chaos (\ref{inscc})
\begin{eqnarray*}\label{Gcid}
   &&\lim\limits_{\epsilon\rightarrow 0}\big\|\epsilon^{s}G_{s}(t)
      \big\|_{\mathfrak{L}^{1}(\mathcal{H}_s)}=0,\quad s\geq2.
\end{eqnarray*}

In case of $s=1$ for series expansion (\ref{mcc}) the following equality is true
\begin{eqnarray*}
   &&\lim\limits_{\epsilon\rightarrow 0}\big\|\epsilon G_{1}(t)-
     g_{1}(t)\big\|_{\mathfrak{L}^{1}(\mathcal{H})}=0,
\end{eqnarray*}
where for arbitrary finite time interval the limit one-particle marginal correlation operator
$g_1(t,1)$ is given by the norm convergent series on the space $\mathfrak{L}^{1}(\mathcal{H})$
\begin{eqnarray}\label{1mco}
   &&\hskip-5mm g_{1}(t,1)=\sum\limits_{n=0}^{\infty}\int\limits_0^tdt_{1}\ldots
      \int\limits_0^{t_{n-1}}dt_{n}\,\mathrm{Tr}_{2,\ldots,n+1}\mathcal{G}^{\ast}_{1}(t-t_{1},1)
      \mathcal{N}^{\ast}_{\mathrm{int}}(1,2)\prod\limits_{j_1=1}^{2}
      \mathcal{G}^{\ast}_{1}(t_{1}-t_{2},j_1)\ldots\\
   &&\hskip+5mm \prod\limits_{i_{n}=1}^{n}\mathcal{G}^{\ast}_{1}(t_{n}-t_{n},i_{n})
      \sum\limits_{k_{n}=1}^{n}\mathcal{N}^{\ast}_{\mathrm{int}}(k_{n},n+1)\prod\limits_{j_n=1}^{n+1}
      \mathcal{G}^{\ast}_{1}(t_{n},j_n)\prod\limits_{i=1}^{n+1}g_1^{0}(i).\nonumber
\end{eqnarray}
In series (\ref{1mco}) the operator $\mathcal{N}^{\ast}_{\mathrm{int}}(j_1,j_2)$ is defined according
to formula (\ref{infOper1}) and the mapping $\mathcal{G}^{\ast}_{1}(t)$ is defined by formula (\ref{grfm}).
For bounded interaction potential series (\ref{1mco}) is norm convergent on the space
$\mathfrak{L}^{1}(\mathcal{H})$ under the condition that: $t<t_0\equiv\big(2\,
\|\Phi\|_{\mathfrak{L}(\mathcal{H}_{2})}\|g_1^{0}\|_{\mathfrak{L}^{1}(\mathcal{H})}\big)^{-1}$.

As a result of the differentiation over the time variable of the operator represented by series
(\ref{1mco}) in the sense of the norm convergence of the space $\mathfrak{L}^{1}(\mathcal{H})$,
we conclude that limit one-particle marginal correlation operator (\ref{1mco}) is governed by
the Cauchy problem of the Vlasov quantum kinetic equation
\begin{eqnarray}\label{Vlasov1}
  &&\frac{\partial}{\partial t}g_{1}(t,1)=\mathcal{N}^{\ast}(1)g_{1}(t,1)+
     \mathrm{Tr}_{2}\,\mathcal{N}^{\ast}_{\mathrm{int}}(1,2)g_{1}(t,1)g_{1}(t,2),\\ \nonumber\\
\label{Vlasov1i}
  &&g_{1}(t)|_{t=0}=g_{1}^0,
\end{eqnarray}
and consequently, in case of pure states we derive the Hartree equation \cite{G12},
i.e. in terms of the kernel $g_{1}(t,q,q')=\psi(t,q)\psi^{\ast}(t,q')$ of the operator
(\ref{1mco}), describing a pure state, equation (\ref{Vlasov1}) reduces to the Hartree equation
\begin{eqnarray*}
    &&i\frac{\partial}{\partial t}\psi(t,q)=-\frac{1}{2}\Delta_{q}\psi(t,q)+
        \int dq'\Phi(q-q')|\psi(q')|^{2}\psi(t,q),
\end{eqnarray*}
where the function $\Phi$ is a two-body interaction potential.

\subsection{On the propagation of initial correlations}
Further, we establish the mean field asymptotic behavior of constructed marginal correlation
operators (\ref{sss}) in case of the initial state specified by the one-particle marginal
density operator with correlations (\ref{insc}).

We assume the existence of a mean field limit of initial one-particle marginal correlation
operator in the sense of equality (\ref{asic1}) and initial correlations as follows:
\begin{eqnarray}\label{asic}
   &&\lim\limits_{\epsilon\rightarrow 0}\big\|g_{n}^{\epsilon}-
      g_{n}\big\|_{\mathfrak{L}^{1}(\mathcal{H}_n)}=0, \quad n\geq2.
\end{eqnarray}
Hence a mean field limit of initial state (\ref{insc}) is specified by the sequence
of limit marginal correlation operators defined on the space
$\mathcal{F}_{\mathcal{H}}={\bigoplus\limits}_{n=0}^{\infty}\mathcal{H}_{n}$:
\begin{eqnarray*}\label{lins}
   &&\hskip-5mm g^{(cc)}=\big(I,g_1^{0}(1),g_{2}(1,2)\prod_{i=1}^{2}g_1^{0}(i),\ldots,
        g_{n}(1,\ldots,n)\prod_{i=1}^{n}g_1^{0}(i),\ldots\big).
\end{eqnarray*}

Under conditions (\ref{asic1}),(\ref{asic}) on initial state (\ref{insc}) there exists
a mean field limit of marginal correlation operators (\ref{sss}) in the following sense
\begin{eqnarray*}\label{asymp}
   &&\lim\limits_{\epsilon\rightarrow 0}\big\|\epsilon^{s}G_{s}(t)-
      g_{s}(t)\big\|_{\mathfrak{L}^{1}(\mathcal{H}_s)}=0, \quad s\geq1,
\end{eqnarray*}
where for $s\geq2$ the limit marginal ($s$-particle) correlation operator $g_s(t)$
is represented by the operator
\begin{eqnarray}\label{dchaos}
     &&\hskip-8mm g_s(t,1,\ldots,s)=\prod _{i_1=1}^{s}\mathcal{G}^{\ast}_{1}(t,i_1)
       g_{s}(1,\ldots,s)\prod_{i_2=1}^{s}(\mathcal{G}_{1}^{\ast})^{-1}(t,i_2)
       \prod\limits_{j=1}^{s}g_{1}(t,j),\quad s\geq2,
\end{eqnarray}
and, respectively, the limit one-particle correlation operator $g_1(t)$ is represented
by the following series expansion
\begin{eqnarray}\label{viterc}
   &&\hskip-8mm g_{1}(t,1)=\sum\limits_{n=0}^{\infty}\,\int\limits_0^tdt_{1}\ldots
        \int\limits_0^{t_{n-1}}dt_{n}\,\mathrm{Tr}_{2,\ldots,n+1}\mathcal{G}^{\ast}_{1}(t-t_{1},1)
        \mathcal{N}^{\ast}_{\mathrm{int}}(1,2)
        \prod\limits_{j_1=1}^{2}\mathcal{G}^{\ast}_{1}(t_{1}-t_{2},j_1)\ldots\\
   &&\hskip+7mm \prod\limits_{i_{n}=1}^{n}\mathcal{G}^{\ast}_{1}(t_{n}-t_{n},i_{n})
        \sum\limits_{k_{n}=1}^{n}\mathcal{N}^{\ast}_{\mathrm{int}}(k_{n},n+1)
        \prod\limits_{j_n=1}^{n+1}\mathcal{G}^{\ast}_{1}(t_{n},j_n)\times\nonumber\\
   &&\hskip+7mm \sum\limits_{\mbox{\scriptsize $\begin{array}{c}\mathrm{P}:(1,\ldots,n+1)=\bigcup_{i}X_{i}\end{array}$}}
        \prod_{X_i\subset \mathrm{P}}g_{|X_i|}(X_i)\prod\limits_{i=1}^{n+1}g_1^0(i).\nonumber
\end{eqnarray}
For bounded interaction potentials series (\ref{viterc}) is norm convergent on the space
$\mathfrak{L}^{1}(\mathcal{H})$ under the condition that:
$t<t_{0}\equiv(2\,\|\Phi\|_{\mathfrak{L}(\mathcal{H}_{2})}\|g_1^0\|_{\mathfrak{L}^{1}(\mathcal{H})})^{-1}$.

The operator $g_{1}(t)$ represented by series (\ref{viterc}) is a solution of the Cauchy
problem of the Vlasov-type quantum kinetic equation with initial correlations:
\begin{eqnarray}\label{Vls}
  &&\hskip-5mm\frac{\partial}{\partial t}g_{1}(t,1)=\mathcal{N}^{\ast}(1)g_{1}(t,1)+\\
  &&\hskip+5mm\mathrm{Tr}_{2}\,\mathcal{N}^{\ast}_{\mathrm{int}}(1,2)
     \prod_{i_1=1}^{2}\mathcal{G}^{\ast}_{1}(t,i_1)(g_{2}(1,2)+1)
     \prod_{i_2=1}^{2}(\mathcal{G}^{\ast}_{1})^{-1}(t,i_2)g_{1}(t,1)g_{1}(t,2),\nonumber\\
     \nonumber\\
\label{Vlasov2c}
  &&\hskip-5mm g_{1}(t)|_{t=0}=g_{1}^0,
\end{eqnarray}
where the operators $\mathcal{N}^{\ast}(1)$ and $\mathcal{N}^{\ast}_{\mathrm{int}}(1,2)$ are
defined according to formula (\ref{infOper1}). We point out that derived kinetic equation
(\ref{Vls}) is non-Markovian quantum kinetic equation.

Thus, mean field dynamics does not create new correlations except of those that generating
by the initial correlations.

The proof of stated results (\ref{dchaos}) and (\ref{viterc}) is based on the validity of
equality (\ref{apg}) for cumulants of asymptotically perturbed groups of operators (\ref{grG})
and the explicit structure of the generating operators of series expansion (\ref{sss}) of
marginal correlation operators, for example, in case of $s=1$ series expansion (\ref{sss})
takes the form
\begin{eqnarray*}\label{skG1}
   &&\hskip-12mm G_{1}(t,1)=\sum\limits_{n=0}^{\infty}\frac{1}{n!}\,\mathrm{Tr}_{2,\ldots,{1+n}}\,\,
      \mathfrak{A}_{1+n}(t)\sum\limits_{\mbox{\scriptsize$\begin{array}{c}\mathrm{P}:(1,\ldots,n+1)=\bigcup_{i}X_{i}\end{array}$}}
      \prod_{X_i\subset \mathrm{P}}g_{|X_i|}^{\epsilon}(X_i)\prod_{i=1}^{n+1}G_{1}^{0,\epsilon}(i),
\end{eqnarray*}
where the generating operator $\mathfrak{A}_{1+n}(t)\equiv\mathfrak{A}_{1+n}(t,1,\ldots,n+1)$
is the $(1+n)th$-order cumulant (\ref{cumcp}) of groups of operators (\ref{grG}).

We remark that the sequence of limit marginal correlation operators (\ref{dchaos}) and (\ref{viterc})
is a solution of the quantum Vlasov hierarchy of nonlinear evolution equations \cite{GP13},
which is described a mean field limit of marginal correlation operators (\ref{sss}) in case
of arbitrary initial states, i.e.
\begin{eqnarray*}\label{gBigfromDFBa_lim}
   &&\hskip-7mm\frac{\partial}{\partial t}g_s(t,1,\ldots,s)=\sum_{i\in (1,\ldots,s)}\mathcal{N}^{\ast}(i)g_{s}(t,1,\ldots,s) +
      \mathrm{Tr}_{s+1}\sum_{i\in(1,\ldots,s)}\mathcal{N}^{\ast}_{\mathrm{int}}(i,s+1)\big(g_{s+1}(t,1,\ldots,\\
   &&\hskip+8mm s+1)+\sum_{\mbox{\scriptsize
      $\begin{array}{c}\mathrm{P}:(1,\ldots,s+1)=X_1\bigcup X_2,\\i\in X_1;s+1\in X_2\end{array}$}}
      g_{|X_1|}(t,X_1)g_{|X_2|}(t,X_2)\big),\quad s\geq1, \nonumber
\end{eqnarray*}
where we used notations similar to accepted in kinetic equation (\ref{Vls}).

We note also that for marginal correlation functionals (\ref{f}) the following equalities hold:
\begin{eqnarray*}\label{cfmf}
    &&\hskip-8mm\lim\limits_{\epsilon\rightarrow 0}\Big\|\epsilon^{s}G_{s}(t,1,\ldots,s\mid G_{1}(t))-\\
    &&\prod _{i_1=1}^{s}\mathcal{G}_{1}^{\ast}(t,i_1)g_{s}(1,\ldots,s)
        \prod_{i_2=1}^{s}(\mathcal{G}_{1}^{\ast})^{-1}(t,i_2)
        \prod\limits_{j=1}^{s}g_{1}(t,j)\Big\|_{\mathfrak{L}^{1}(\mathcal{H}_s)}=0,\quad s\geq2,\nonumber
\end{eqnarray*}
where the limit one-particle (marginal) correlation operator $g_{1}(t)$ is represented
by series expansion (\ref{viterc}), i.e. it is a solution of the Cauchy problem (\ref{Vls}),(\ref{Vlasov2c}).


\section{Conclusion}

The marginal correlation operators (\ref{sss}) give an equivalent approach to the description
of the evolution of states of large particle quantum systems in comparison with the marginal
density operators. The macroscopic characteristics of fluctuations of observables are directly
determined by marginal correlation operators on the microscopic scale \cite{BQ},\cite{GP13}.

This paper deals with a quantum system of a non-fixed, i.e. arbitrary but finite, number of
identical (spinless) particles obeying Maxwell -- Boltzmann statistics. The obtained results
can be extended to large particle quantum systems of bosons and fermions \cite{GP11}.

In the paper it was established that a nonperturbative solution of the Cauchy problem of the
nonlinear quantum BBGKY hierarchy (\ref{gBigfromDFBa}),(\ref{gBigfromDFBai}) for a sequence
of marginal correlation operators is represented in the form of series expansion (\ref{sss})
over particle subsystems which generating operators are corresponding-order cumulant (\ref{cc})
of groups of nonlinear operators (\ref{rozvNh}). In case of initial states specified by
a sequence of the marginal correlation operators that satisfy a chaos property (\ref{inscc})
the correlations generated by dynamics of large particle quantum systems (\ref{mcc}) are
completely determined by cumulants (\ref{cumulantP}) of groups of operators (\ref{grG})
of the von Neumann equations.

The concept of quantum kinetic equations in case of initial states specified in terms of
a one-particle density operator and correlation operators (\ref{gkec}), for instance, the
correlation operators, characterizing the condensed states, was considered. We remark that
in case of pure states the quantum Vlasov-type kinetic equation with initial correlations
(\ref{Vls}) can be reduced to the Gross -- Pitaevskii-type kinetic equation.

We also emphasize that the natural Banach spaces for the description of states of large
particle quantum systems, for instance, the operator spaces containing the sequence of operators 
of equilibrium state, are different from the used Banach space \cite{G12}. In paper \cite{YG} 
it was introduced the space of sequences of bounded translation invariant operators, making 
it a better choice for the description of quantum correlations.

\end{document}